\documentclass[reprint,prl,twocolumn,superscriptaddress,showpacs,showkeys,floatfix,preprintnumbers]
{revtex4}

\usepackage{graphicx}
\usepackage{slashed}
\usepackage{amsmath,amsthm}
\usepackage{subfigure}
\usepackage[multidot]{grffile} 
\usepackage{color}
\usepackage{bbold}

\def\beq{\begin{equation}}
\def\eeq{\end{equation}}
\def\bea{\begin{eqnarray}}
\def\eea{\end{eqnarray}}
\def\bq{\begin{quote}}
\def\eq{\end{quote}}
\def\ben{\begin{enumerate}}
\def\een{\end{enumerate}}
\def\bit{\begin{itemize}}
\def\eit{\end{itemize}}
\def\nn{\nonumber}

\def\Id{\mathbb{1}}

\begin{document}

\title{Thermal evolution of the Schwinger model with Matrix Product Operators}

\date{\today}

\author{M. C. Ba\~nuls}
\affiliation{Max-Planck-Institut f\"ur Quantenoptik, Hans-Kopfermann-Str. 1, 85748 Garching, Germany}

\author{K. Cichy}
\affiliation{Goethe-Universit\"at, Institut f\"ur Theoretische Physik,
Max-von-Laue-Stra\ss e 1, D-60438 Frankfurt a.M., Germany}
\affiliation{NIC, DESY, Platanenallee 6, D-15738 Zeuthen, Germany}
\affiliation{Adam Mickiewicz University, Faculty of Physics, Umultowska 85, 61-614 Poznan, Poland}

\author{J. I. Cirac}
\affiliation{Max-Planck-Institut f\"ur Quantenoptik, Hans-Kopfermann-Str. 1, 85748 Garching, Germany}

\author{K. Jansen}
\affiliation{NIC, DESY, Platanenallee 6, D-15738 Zeuthen, Germany}

\author{H. Saito}
\affiliation{NIC, DESY, Platanenallee 6, D-15738 Zeuthen, Germany}

\noaffiliation

\begin{abstract}
We demonstrate 
the suitability of tensor network techniques for describing 
 the thermal evolution of lattice gauge theories. 
As a benchmark case, we have studied the temperature dependence of the chiral 
condensate in the Schwinger model, using matrix product operators to 
approximate the thermal equilibrium states for finite system sizes with non-zero lattice spacings.
We show how these techniques allow for 
reliable extrapolations in bond dimension, step width, system size
and lattice spacing,
and for a systematic estimation and control of all error sources involved in the calculation.
The reached values of the lattice spacing are small enough to capture the 
most challenging region of high temperatures
and the final results are consistent with the 
analytical prediction by Sachs and Wipf over a broad
temperature range. 
\end{abstract}



\preprint{DESY 15-060, SFB/CPP-14-125}

\maketitle

\section{Introduction}

Tensor network (TN) techniques have recently revealed 
their potential to study lattice gauge theories (LGT).
Numerical studies have demonstrated that 
matrix product states (MPS) can accurately describe
ground state and low excited levels of the 
Schwinger model \cite{Byrnes:2002nv,Cichy:2012rw,Banuls:2013jaa,Banuls:2013zva,Buyens:2014pga,Buyens:2013yza}
and of related quantum link models \cite{Rico:2013qya,Silvi:2014pta},
and tensor renormalization group methods have been used to 
evaluate the path integral \cite{shimizu2014grassmann,Shimizu:2014fsa}. 
More generally, a framework has been proposed to construct 
gauge invariant TN states in higher dimensions
\cite{Tagliacozzo:2014bta,haegeman15gauging}.

The TN ansatz can also be used to describe thermal equilibrium states.
Non-zero temperature studies have played a major role
in lattice QCD computations, establishing a crossover behavior
of QCD in the early universe \cite{Borsanyi:2012ve}. Such 
calculations will be very important in the future for understanding 
QCD matter, see \cite{Heinz:2015tua}. 
Different from most LGT calculations, tensor network methods 
work in the Hamiltonian formalism.
In this approach, it is in principle possible to follow 
the complete thermal evolution of the system. 
Thus, employing Hamiltonian calculations could allow us to map out the temperature 
dependence of many physical quantities over a much broader temperature 
regime and in a more precise way than 
conventional Markov Chain Monte Carlo methods. 

In this work, we show that the matrix product operator (MPO) ansatz 
can accurately describe the thermal equilibrium states of the Schwinger 
model.
To this end, we investigate the temperature dependence of the chiral 
condensate in the continuum limit for the massless case, for which analytical
results were provided in \cite{Sachs:1991en}. 
Our results are consistent with the analytical prediction 
at all (dimensionless) inverse temperatures $g \beta\in[0,6]$.
Our lattice calculation using MPO requires a series of consecutive extrapolations. 
We describe how to carry out these steps and demonstrate that all systematic errors 
inherent to the method can be controlled and systematically improved.
Thus, the procedure yields reliable continuum values
and is applicable also when the exact value is completely unknown.

We use Gauss' law to integrate out the gauge degrees of freedom and
apply TN states to describe the fermionic degrees of freedom in the exact physical subspace,
as in \cite{Banuls:2013jaa,Banuls:2013zva}.
Here we demonstrate that this approach, initially presented in \cite{Saito:2014bda}, 
is also suitable for thermal states.
Alternatively, one could include both fermionic and bosonic degrees of freedom and
impose gauge symmetry on the tensors, as in
\cite{Byrnes:2002nv,Buyens:2014pga,Buyens:2013yza,Rico:2013qya,Silvi:2014pta}, 
 but in that case the gauge degrees of freedom need to be truncated, which 
 introduces an additional extrapolation in the procedure, which also has to be taken into account in the
systematic errors.

\section{The model and the Hamiltonian setup}
\label{sec:model}

The Schwinger model \cite{schwinger62} or QED in $1+1$ dimensions, is frequently used 
as a testbench for lattice calculations.
In order to apply TN methods, we work in the Hamiltonian formalism (see e.g.~\cite{Banuls:2013jaa}), 
which implies we have to impose Gauss' law as a constraint on physical states.
The Kogut-Susskind Hamiltonian \cite{kogut75sc} can be mapped to a spin system by a 
Jordan-Wigner transformation \cite{Banks:1975gq}.
Using Gauss' law, the gauge degrees of freedom can be eliminated \cite{hamer97free},
since the electric flux on a link is completely determined by the spin content and
the background field, so that the model can be written:
\begin{align}
H=& x\sum_{n=0}^{N-2} \left [ \sigma_n^+\sigma_{n+1}^- +  \sigma_n^-\sigma_{n+1}^+ \right ]+\frac{\mu}{2}\sum_{n=0}^{N-1} \left [ 1 + (-1)^n \sigma_n^z \right ]
\nonumber \\
&+\sum_{n=0}^{N-2} \left [ \ell +\frac{1}{2}\sum_{k=0}^n ((-1)^k+\sigma_k^{z})\right ]^2,
\label{eq:H}
\end{align}
where $\ell$ is the boundary electric field (on the leftmost link), 
which can describe the background field, and the parameters of the
model are $x=\frac{1}{g^2a^2}$, $\mu=\frac{2 m}{g^2 a}$, in terms of the lattice spacing, $a$, 
the fermion mass, $m$, and the coupling, $g$.

It is then possible to use a basis 
$|\ell\rangle|i_0 i_1 \ldots i_{N-1}\rangle$ to describe
the physical space.
For finite systems, the value of $\ell$ is conserved. In the following, 
we will consider the case $\ell=0$ and omit it from the basis.
We focus on the temperature dependence of the 
 chiral condensate, $\Sigma=\langle \bar{\Psi}\Psi \rangle /g$,
which is the 
 order parameter of the chiral symmetry breaking, and which in terms of 
spin operators reads
$\Sigma=\frac{\sqrt{x}}{N}\sum_n (-1)^n \frac{1+\sigma_n^z}{2}$.

\section{The ansatz}
\label{sec:mpo}

An MPS for a system of $N$ sites with internal dimension $d$ and individual basis
$\{|i\rangle\}_{i=1}^{d}$ is a state of the form
$ |\Psi\rangle 
 =\sum_{i_0,\ldots
  i_{N-1}=1}^d \mathrm{tr}(A_0^{i_0}\ldots A_{N-1}^{i_{N-1}}) |i_0,\ldots i_{N-1}\rangle,$
 where each $A_k^i$ is a $D$-dimensional matrix, and the bond dimension, 
 $D$, determines the number of free parameters in the ansatz \cite{verstraete04dmrg,vidal03eff,perez07mps}.
MPS are known to provide good approximations to ground states of local
Hamiltonians in the gapped phase \cite{hastings07area}
and have also been successfully used for more general situations \cite{schollwoeck11age}. 
The analogous ansatz in the space of operators \cite{verstraete04mpdo,zwolak04mpo,pirvu10mpo}
is called matrix product operators (MPO),
and can be used to efficiently approximate thermal states of local Hamiltonians
\cite{hastings06thermal,molnar2015thermal}.

To find an MPO approximation to the Gibbs state, $\rho \propto e^{-\beta H}$,
 a Suzuki-Trotter decomposition is applied to the exponential, 
with the Hamiltonian split into several terms whose exponentials are easily written or approximated as MPO.
In the case of Hamiltonian \eqref{eq:H}, it is convenient to split $H=H_e+H_o+H_z$,
where $H_{e(o)}$ contains the $\sigma_n^+ \sigma_{n+1}^- + h.c.$ terms for even (odd) $n$, and $H_z$
contains the mass terms and the long range $\sigma_n^z \sigma_m^z $ interactions.
The exponentials of $H_{e(o)}$ can be easily written as exact MPO \cite{pirvu10mpo}. For $H_z$, instead, 
the exponential can only be approximated.
Adopting a 2nd-order Trotter expansion and, for $H_z$, a 1st-order Taylor expansion, we can
write:
\beq
   \rho(\beta)
   \approx \left[ e^{- \frac{\delta}{2} H_e} \left( 1-\frac{\delta}{2} H_z \right) e^{-\delta H_o} 
      \left( 1-\frac{\delta}{2} H_z \right) e^{-\frac{\delta}{2} H_e} \right]^M,  
\label{eq:e^deltaH}
\eeq
where $\delta=\beta/M$ is the step width, and the final error for fixed $\beta$ will be $\cal{O}(\delta)$, dominated 
by the Taylor expansion.

Starting from the identity operator, which corresponds to infinite temperature, $\rho(\beta=0)$,
we apply successive Euclidean evolution steps. After each of them a truncation is carried out to find an
MPO approximation to the result. 
To this end, using a Choi isomorphism, the MPO is mapped to an MPS with local physical dimension $d^2$,
 and an alternating least squares procedure is applied to minimize 
 the Euclidean distance between the vectorized MPO for the new and evolved states.
Since the truncation does not preserve the positivity of the whole state, it is more convenient to compute
$\rho(\beta/2)^{\dagger} \rho(\beta/2)$, where the Trotter expansion explained above is used for each 
factor. \footnote{Notice that this is equivalent to the purification ansatz.}

\section{Results}
\label{sec:results}
To compute the chiral condensate in the infinite volume and 
continuum limits,
we approximate the thermal state at each $g\beta\in[0,6]$ 
over a range of values of $x\in[4,65]$. For each value, we consider various system sizes, with
$N/\sqrt{x}\in [16,24]$ to ensure consistent physical volumes, 
and for each of them, different step widths, $\delta$, and bond dimensions.
We thus need to control effects of successively extrapolating in $D$, $\delta$, $N$ and $x$.

\begin{figure}[floatfix]
\includegraphics[width=.425\columnwidth]{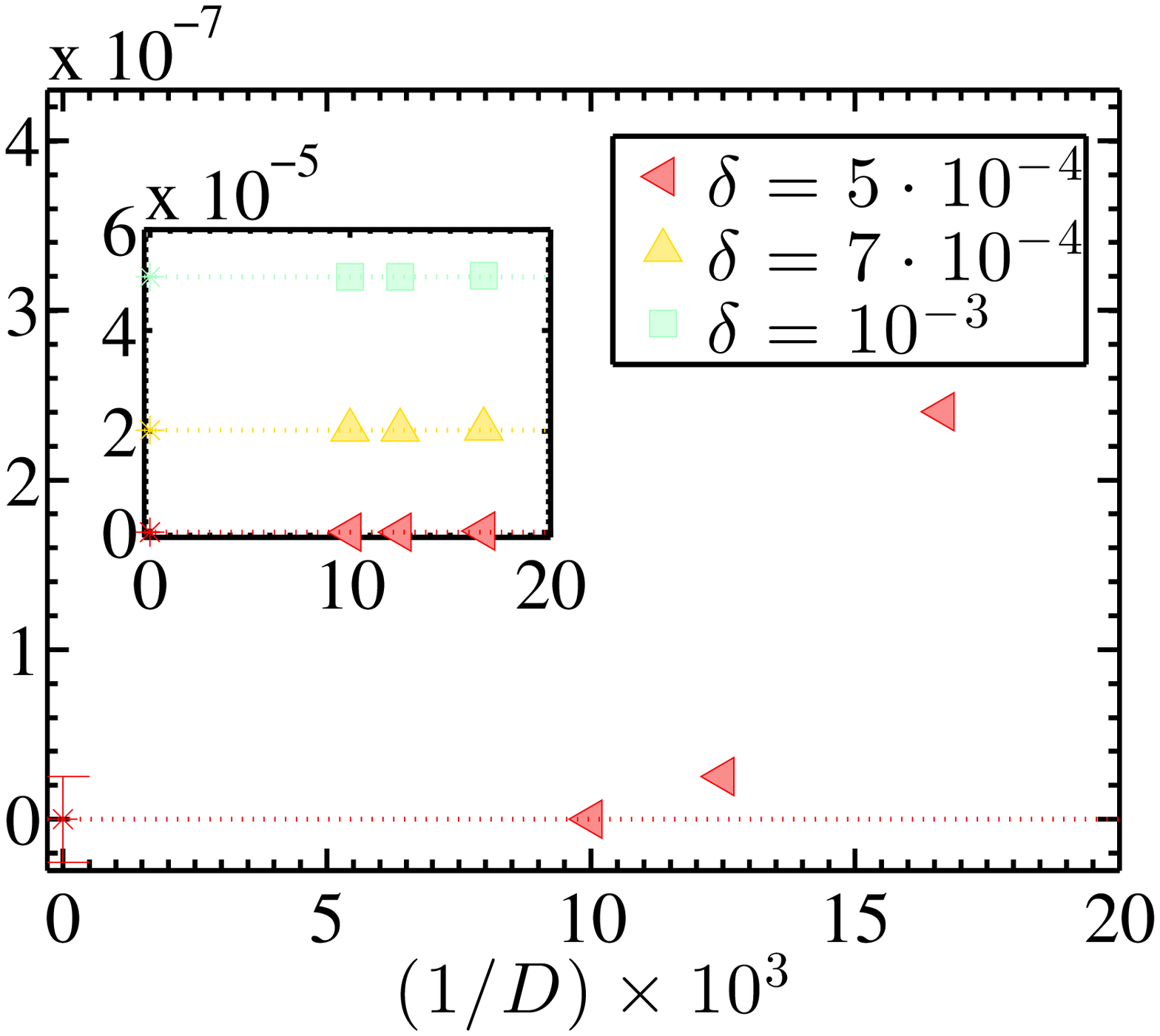}
\hspace{.05\columnwidth}
\includegraphics[width=.47\columnwidth]{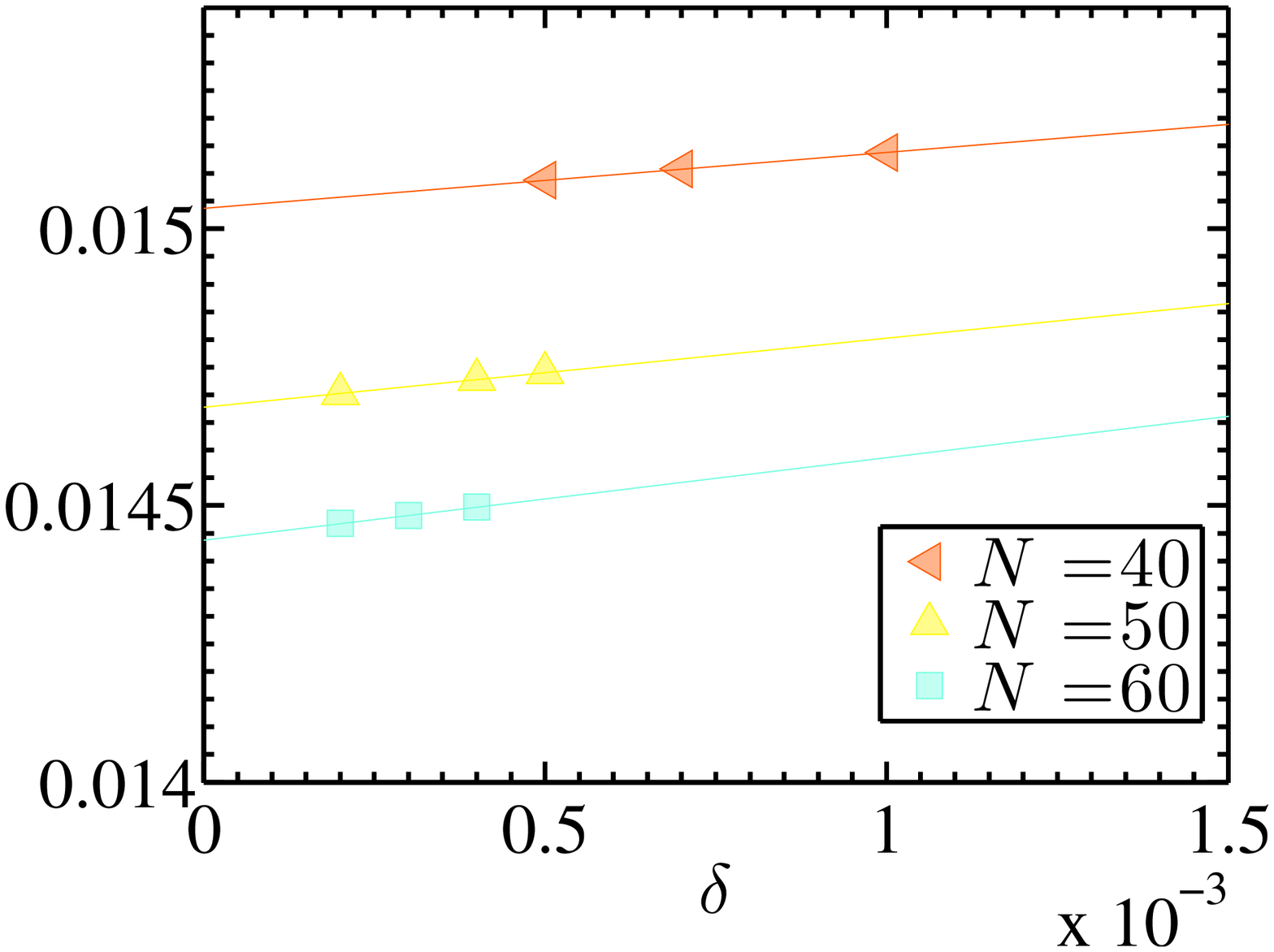}
\caption{Left: convergence of condensate value with bond dimension, $D$, for $g\beta=0.4$, $x=6.25$, 
$N=40$ and $\delta=5\cdot 10^{-4}$. Shown is the deviation with respect to the final value, 
$\Sigma=0.01508769$. Looking at various $\delta$ (inset) we find that the truncation error is much smaller than the variation due to the finite step width.
 Right: linear $\delta$ extrapolation for the same $x$ and $g\beta$ and several system sizes. }
\label{fig:Danddelta}
\end{figure}

The limited bond dimension, $D$, used for each fixed set of values 
($x$, $N$, $\delta$) induces a systematic truncation error.
The MPS family being complete, the results converge to the exact value
for the given problem in the limit of very large $D$ (of the order of the 
dimension of the operator space) \cite{perez07mps}.
From our finite $D$ results, we estimate the final value of $\Sigma$ as the one
obtained for the largest achieved bond dimension, $D_{\mathrm{max}}$, and the error
as the difference between this value and the one obtained 
from $D_{\mathrm{max}}-20$, as illustrated in the left panel of Fig.~\ref{fig:Danddelta}.

A second source of systematic error is the finite step width, $\delta$.
Although we use a second order Suzuki-Trotter expansion, the Taylor
approximation in \eqref{eq:e^deltaH} induces linear corrections, $\cal{O}(\delta)$.
We can thus extrapolate linearly to obtain the value as $\delta\to0$,
 as illustrated by the right panel in Fig.~\ref{fig:Danddelta} for selected examples.
Furthermore,  the Taylor approximation 
requires that the value of $\delta$ employed in 
the calculation is below a certain threshold, to ensure 
convergence of the expansion.
We find that values $\delta=10^{-6}$-$10^{-3}$ are sufficiently small for
the considered values of $x$ and $N$.

\begin{figure}[floatfix]
\includegraphics[width=.46\columnwidth]{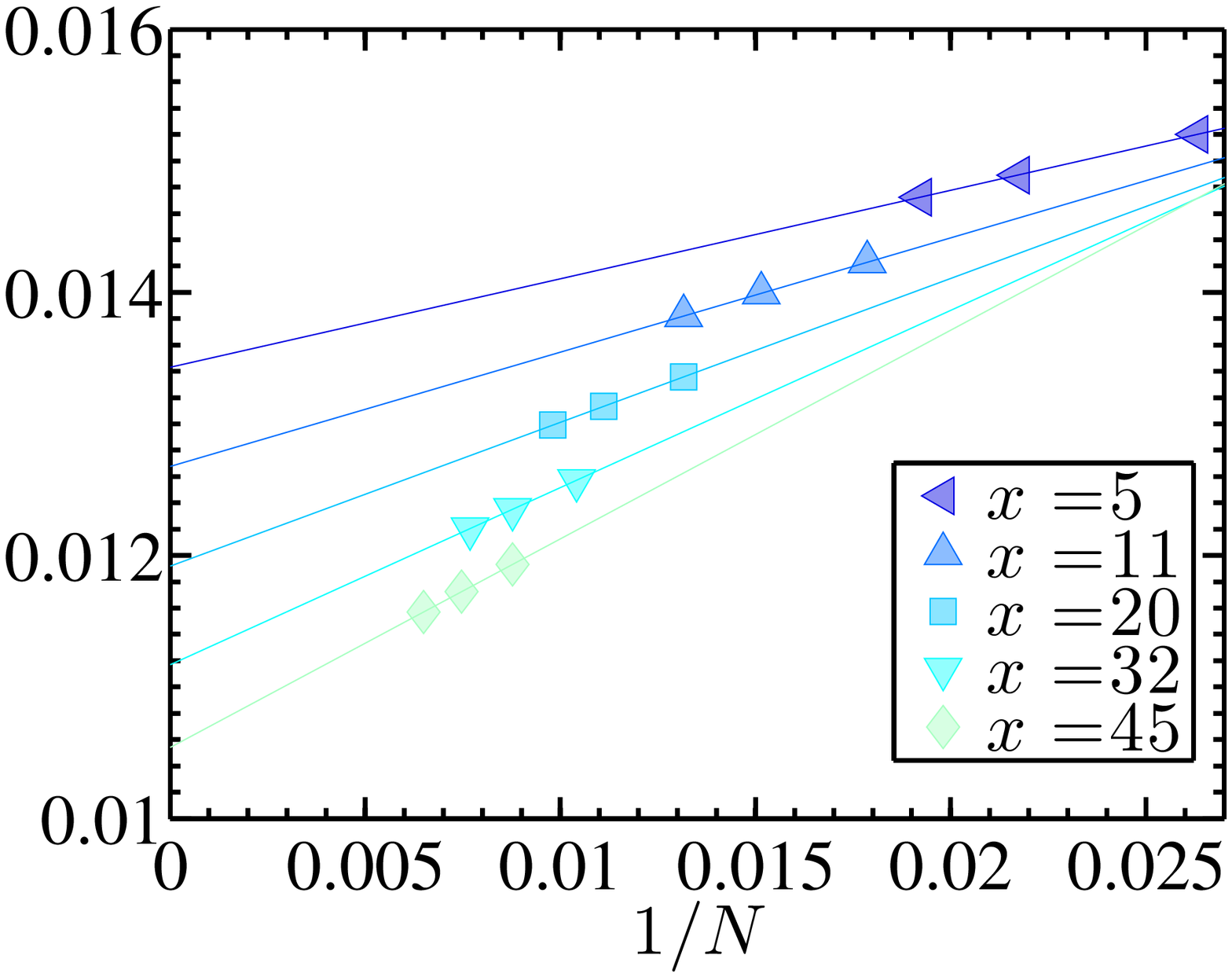}
\hspace{.05\columnwidth}
\includegraphics[width=.45\columnwidth]{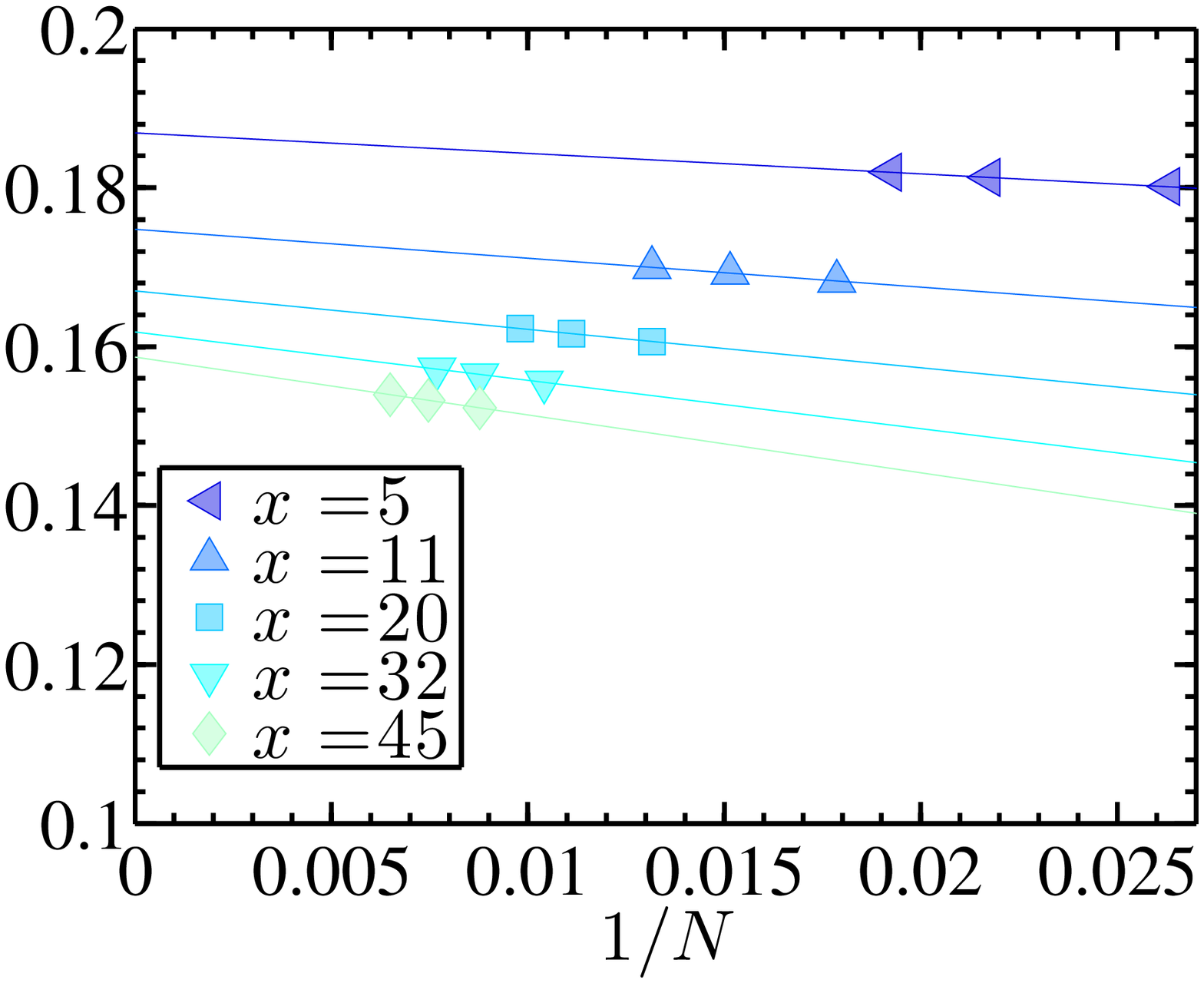}
\caption{Infinite volume extrapolation of the condensate values for 
several values of $x$ at
$g\beta=0.4$ (left) and $4.0$ (right).}
\label{fig:Nandx}
\end{figure}

The previous steps yield a result for each pair ($x$, $N$).
As in \cite{Banuls:2013jaa,Banuls:2013zva}, we then 
find the thermodynamic limit by fitting the results to a linear function in $1/N$.
The left panel of Fig.~\ref{fig:Nandx} shows how accurately
this extrapolation fits our results for the considered values of $x$.

\begin{figure}[floatfix]
\includegraphics[width=.46\columnwidth]{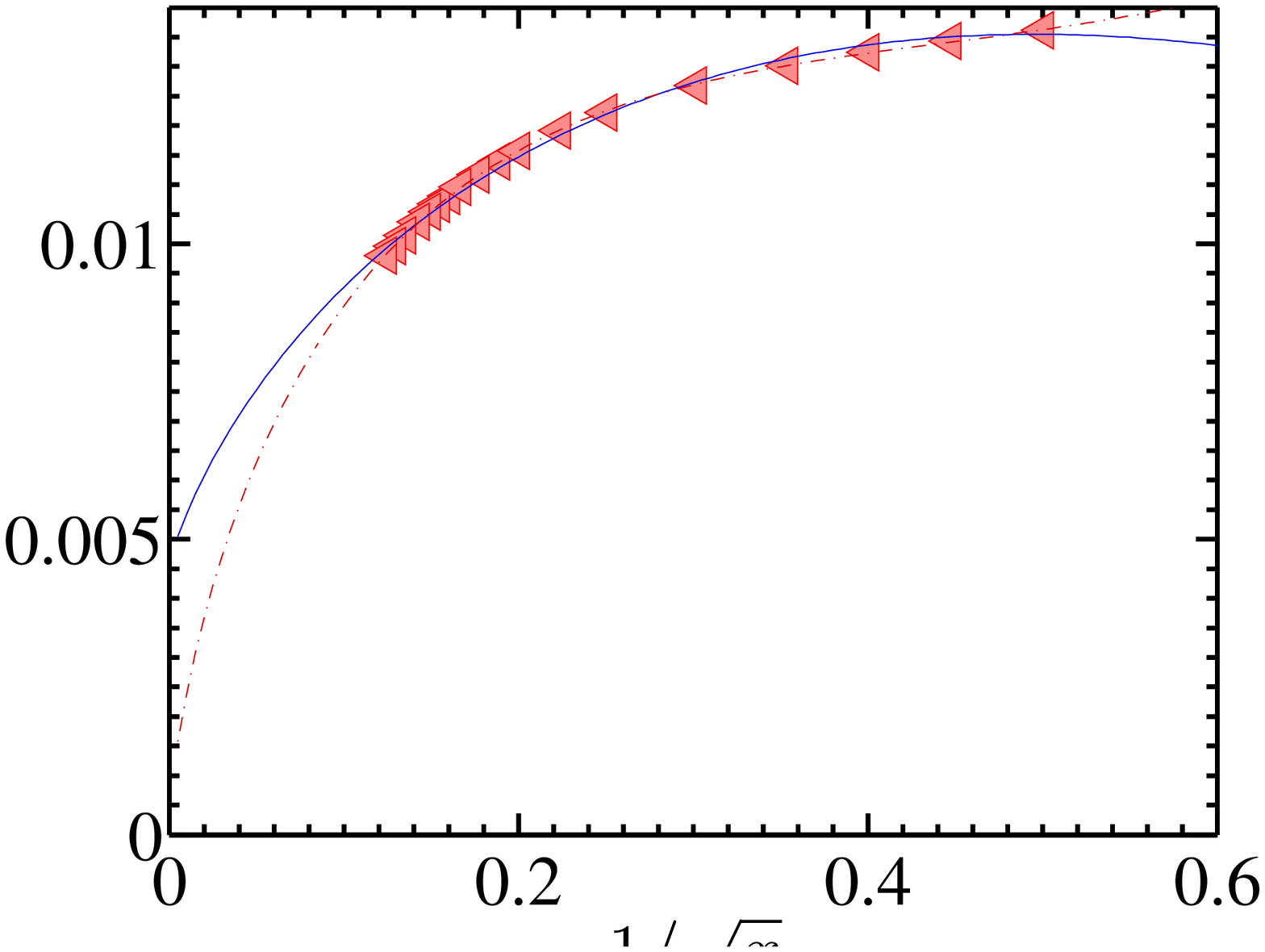}
\hspace{.05\columnwidth}
\includegraphics[width=.45\columnwidth]{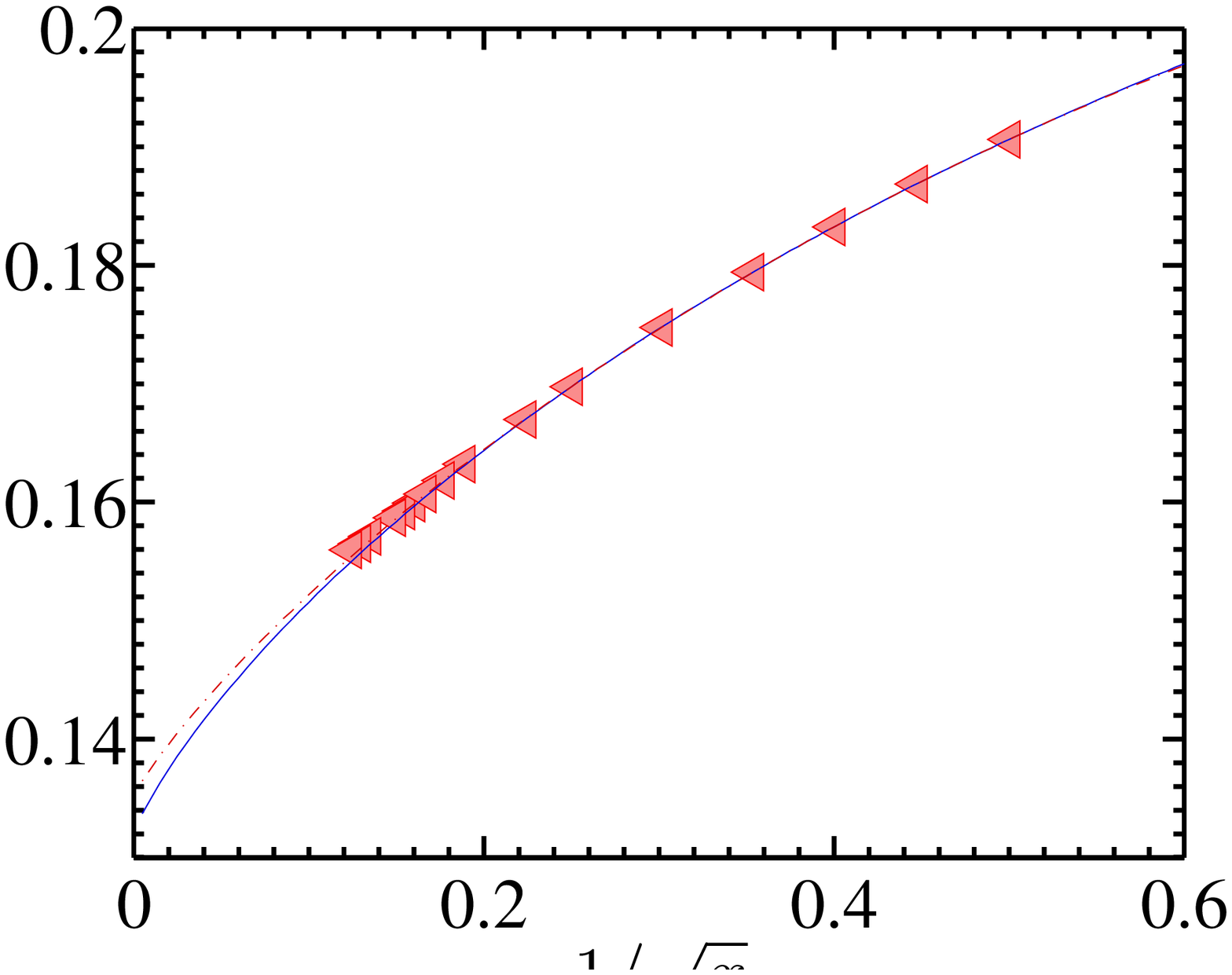}
\caption{Condensate, $\Sigma$, as a function of the lattice spacing for $g\beta=0.4$ (left)
and $4.0$ (right). The lines show the continuum extrapolation using $f_1$ (solid blue) and $f_2$ (dash-dotted red).}
\label{fig:x}
\end{figure}

From the infinite volume results for each lattice spacing at fixed $g\beta$,
we can perform the continuum extrapolation.
As we showed in \cite{Banuls:2013zva}, the condensate exhibits 
logarithmic corrections $\mathcal{O}(a\log(a))$. 
Hence we try two fitting functions, which additionally include linear or linear and quadratic corrections in $a$,
\begin{align}
f_1(x)&=\Sigma_{\rm cont}
+ \frac{a_1}{\sqrt{x}}\log(x)+\frac{b_1}{\sqrt{x}} , 
\nn
\\
f_2(x)&=\Sigma_{\rm cont} 
+ \frac{a_2}{\sqrt{x}}\log(x)+\frac{b_2}{\sqrt{x}}+\frac{c_2}{x}. 
\label{eq:f1f2}
\end{align}
To include the uncertainty from the choice of the functional form of the fitting ansatz, we finally take
as central value the result from the fit to $f_1$, and as systematic error
the difference between both.
Fig.~\ref{fig:x} demonstrates these fits for two very different temperatures, $g\beta=0.4$ and $4.0$,
and shows that quadratic corrections will only be significant at lower temperatures.
Higher order corrections do not provide any significant improvement 
for the fitting range $x\in[4,65]$.

\begin{figure}[floatfix]
\includegraphics[width=.45\columnwidth]{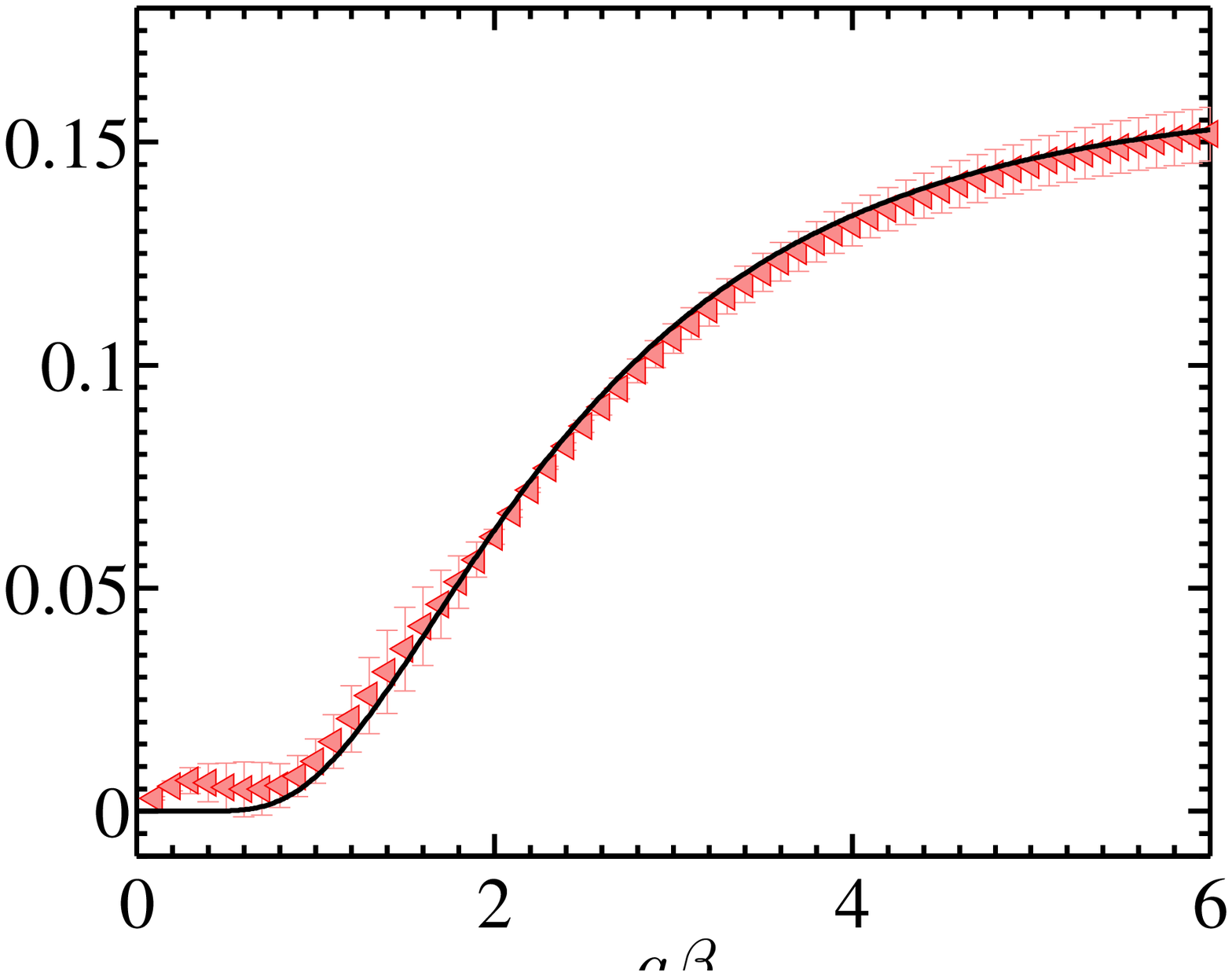}
\hspace{.05\columnwidth}
\includegraphics[width=.45\columnwidth]{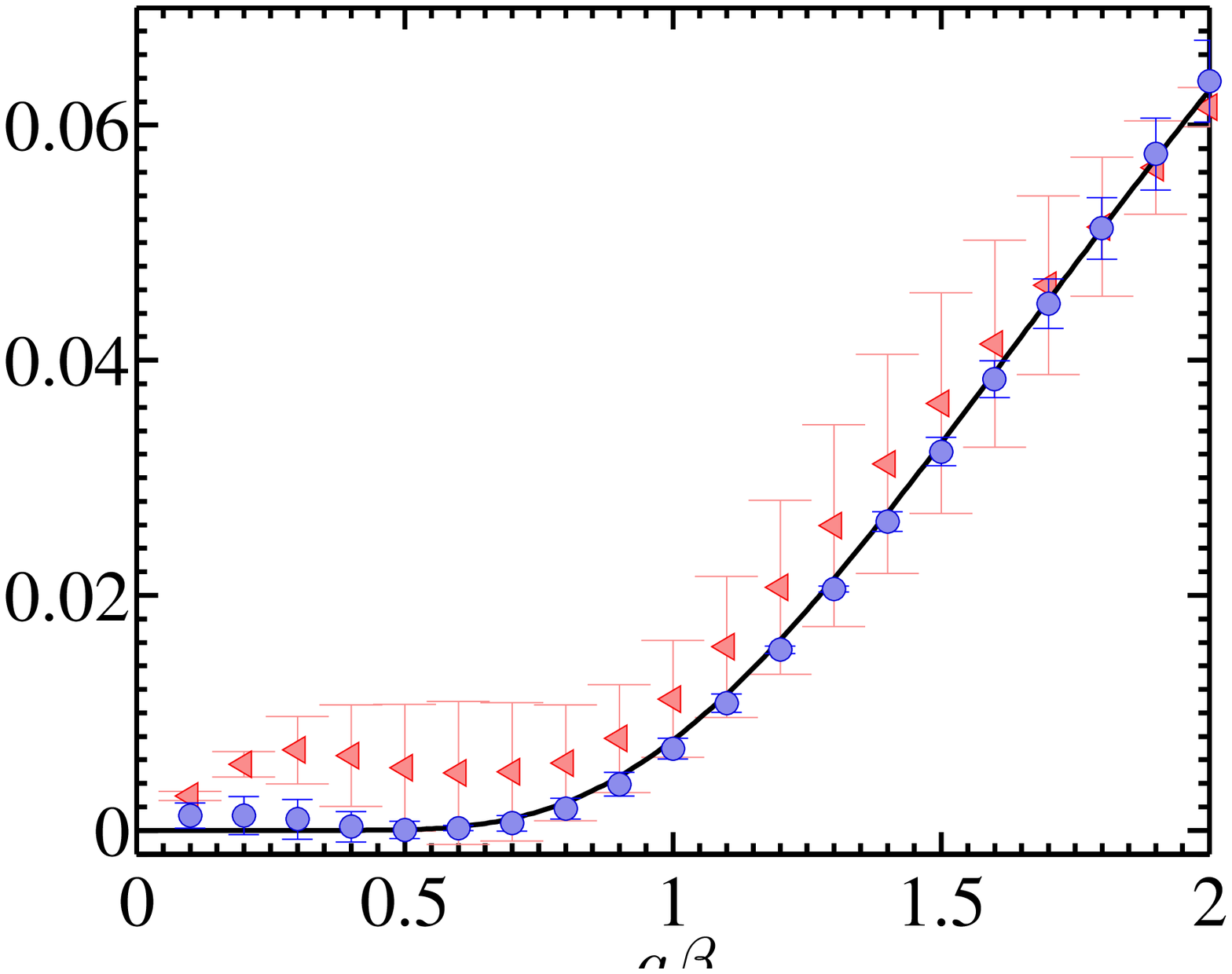}
\caption{
Temperature dependence of the chiral condensate from data with $x\leq65$ and exact 
gauge sector (red triangles), compared to \cite{Sachs:1991en} (solid line). 
For the lowest $g\beta$ (right), the results deviate from the exact ones. Using the $L_{\mathrm{cut}}=10$
truncation, we reach smaller lattice spacings and recover the consistency with the analytical results
(blue circles).
}
\label{fig:final}
\end{figure}

After performing for each value of $g\beta$ all the steps described above, 
we obtain for the chiral condensate the temperature
dependence shown in Fig.~\ref{fig:final}. 
Comparing to the analytical result in \cite{Sachs:1991en},
we find excellent agreement for all $g\beta\geq 0.5$.
Although the central values lie very close to the exact results, the errors shown in
Fig.~\ref{fig:final} 
seem relatively large because they include the propagated errors from the extrapolations in $D$,
$\delta$, $N$ and $x$,
as well as the systematic error from the form of the fitting ansatz for the continuum extrapolation. 
Our approach makes it possible to fully control all sources of uncertainties. 
This rigorous account of errors is crucial to ensure that the technique can be used 
in a general situation, for which no analytical results are available.

Different kinds of errors contribute distinctly at different temperatures. For small $g\beta$,
cut-off effects are enhanced, and systematic errors from the choice of the fitting ansatz
can be an order of magnitude larger than other errors.
Lowering the temperature, the effect becomes smaller, while other errors grow, in particular 
the propagated error from the $D\rightarrow\infty$ extrapolation.
Interestingly, at intermediate values, $g\beta\approx 2$, the slope of the
$N\rightarrow\infty$ extrapolations changes sign (see Fig.~\ref{fig:Nandx}).
Moreover, for this region of temperatures, 
the continuum limit extrapolations from  $f_1$ and $f_2$
are very close.
For larger $g\beta$, all errors grow, 
but the increase is faster for $D$ and $\delta$ extrapolations 
so that at the end of our $g\beta$ range
they are an order of magnitude larger 
than the systematic error from the choice of the fitting ansatz.

The approach is systematically improvable, because it is clear how to reduce each uncertainty.
For the present analysis, the relatively most important error comes from the extrapolation in $D$. 
We chose a rather conservative criterion to estimate this error and it is natural to
expect that much more accurate results can be obtained by checking the convergence in bond dimension
with larger values of $D$.
The cost of the computation scales with $D^3$, which makes the scan over a rather broad range of $D$
feasible.
It is nevertheless remarkable that the very accurate results presented here were obtained with a 
relatively small $D\leq 100$. This shows how adequate the MPO ansatz is for thermal states
in this gauge theory.

For the region of very small $g \beta$, we find a significant deviation from the analytical results (see
Fig.~\ref{fig:final}),
although the individual points at finite $x$ are accurate enough,
because much smaller lattice spacings 
are required in order to correctly capture the asymptotic behavior. 
Using the procedure described above, it is possible to reach larger values of $x$,
by incurring a higher computational cost, since
 the required system size grows as $\sqrt{x}$ to maintain a 
consistent physical volume,
and correspondingly the threshold $\delta$ that ensures convergence decreases.

Using an alternative approximation 
for the exponential of $H_z$ that avoids the Taylor expansion,
it is however possible to explore the region of larger $x$ at a lower computational cost.
Indeed, $e^{-\delta H_z}$ can be written exactly as an 
MPO of bond dimension ${\cal O}(N)$.
For systems of several hundreds of sites this is unpractical, but
this exponential can be approximated
by projecting out those spin configurations that correspond to an electric flux larger than a
certain cutoff, $L_{\mathrm{cut}}$, on any of the links.
This results in an MPO with bond dimension $2 L_{\mathrm{cut}} +1$.
Notice that this truncation is equivalent to limiting the maximum occupation number for the 
bosonic degrees of freedom, as done in \cite{Byrnes:2002nv,Buyens:2013yza,Buyens:2014pga}
for pure states.
However, the latter results in twice as large system sizes, since bosonic degrees of freedom are kept explicitly,
and, in principle, in local dimensions that scale as $(2 L_{\mathrm{cut}}+1)^2$ for the additional sites in the 
MPO.~\footnote{By using symmetric tensors that enforce Gauss' law, the squaring of local bosonic
dimensions can nevertheless be avoided~\cite{boyePriv}.}

\begin{figure}[floatfix]
\includegraphics[width=.45\columnwidth]{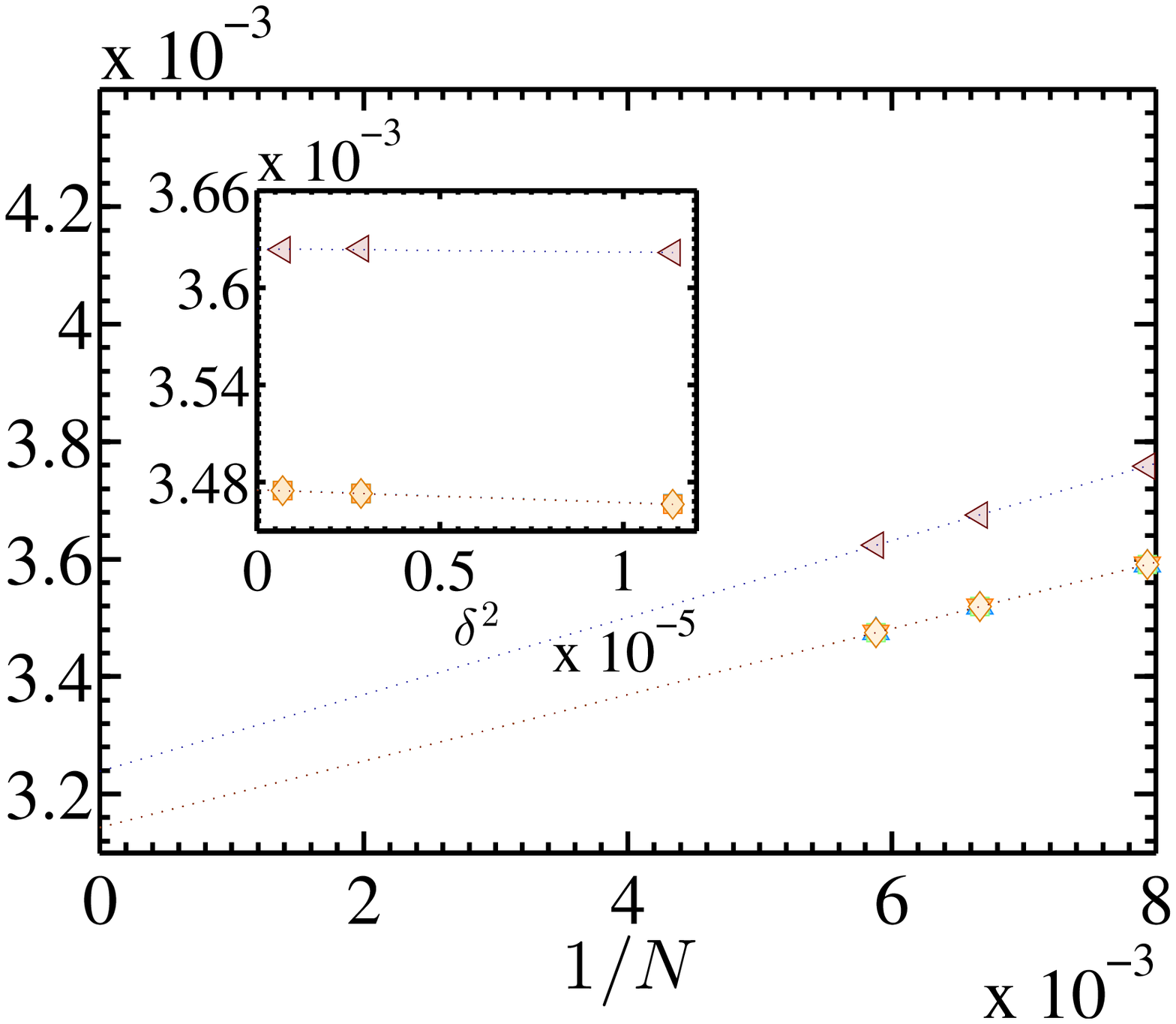}
\hspace{.05\columnwidth}
\includegraphics[width=.45\columnwidth]{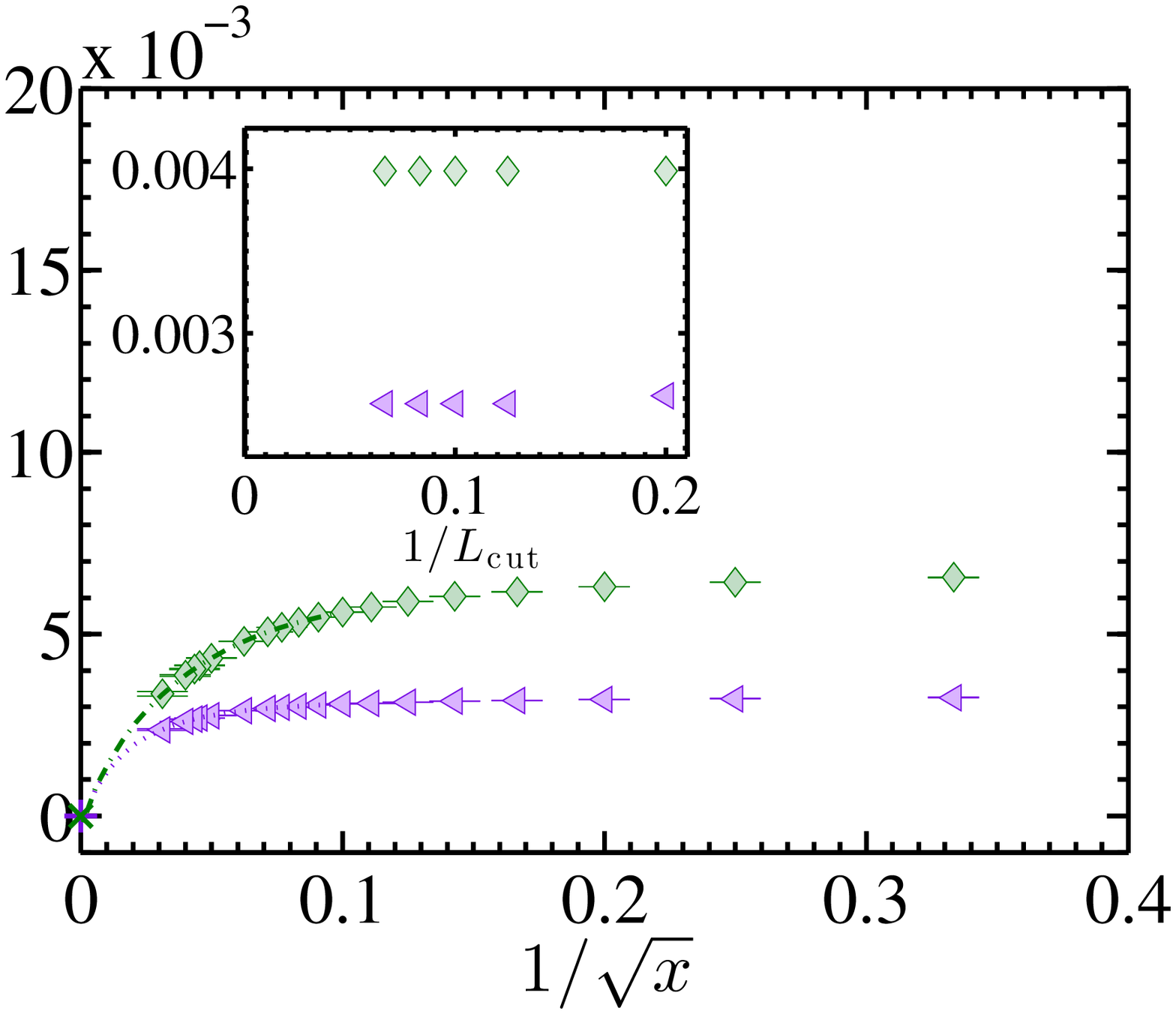}
\caption{Condensate values with a truncation $L_{\mathrm{cut}}$ of the maximum electric flux per link.
Left: finite volume extrapolation for $x=55$, $g\beta=0.1$ from $L_{\mathrm{cut}}=5$ (left pointing red
triangles), $8$, $10$, $12$ (different shapes and colors, indistinguishable in the plot) and $15$ (orange
diamonds). The inset shows the extrapolation in Trotter parameter, $\delta$, for $N=170$ and
$L_{\mathrm{cut}}=8-15$. 
Right: continuum limit for $g\beta=0.1$ (left pointing purple triangles) and $0.2$
(green diamonds). The corresponding exact results from~\cite{Sachs:1991en} are indicated on the vertical
axis. The inset shows explicitly the dependence on the cutoff for these $g\beta$ values at  $x=55$.
}
\label{fig:truncation}
\end{figure}

Although the cost of applying the MPO for the projected exponential is 
higher than that of the Taylor approximation,
the step width error is now ${\cal O}(\delta^2)$, determined
by the second order Trotter expansion, and there is no threshold value
for $\delta$, which allows us to reach the same $g\beta$ with fewer steps.
To explore the small $g\beta$ region, we study the range $x \in [4,1024]$. For each value of $x$, 
we compute different system sizes, step widths and $L_{\mathrm{cut}}$ values, 
and for each of them bond dimensions up to $D=160$.
As described above, we successively extrapolate $1/D\to 0$ (as before), $\delta \to 0$ (linear in $\delta^2$)
and $1/N\to 0$, as illustrated by Fig.~\ref{fig:truncation} (left).
To account for the additional systematic error due to the cutoff parameter, we can also extrapolate in
$L_{\mathrm{cut}}$.
However, comparing the results for $L_{\mathrm{cut}}\in[5,15]$,
we observe (see Fig.~\ref{fig:truncation})
that the effect is very small, and results for $L_{\mathrm{cut}}\geq8$
are compatible within our numerical precision (inset of Fig.~\ref{fig:truncation} right), so here
we present the
results for $L_{\mathrm{cut}}=10$ and leave the detailed analysis of the cutoff effects to a more
technical work~\cite{inprep}.
 
This relatively small $L_{\mathrm{cut}}$ allows us to study the lattice effects for the smallest $g\beta$.
As shown in Fig.~\ref{fig:truncation}, $g\beta=0.1$ requires  $x \approx 300$ or larger
to obtain an accurate continuum extrapolation.
We observe that higher order corrections are present and adopt as central value the result of the fit
$f_3(x)=\Sigma_{\rm cont} 
+ \frac{a_3}{\sqrt{x}}\log(x)+\frac{b_3}{\sqrt{x}}+\frac{c_3}{x}+\frac{d_3}{x \sqrt{x}}$
\footnote{For the smallest $g\beta\leq 0.5$, we use $x\geq100$, for $0.5 < g\beta \leq 1.5$, $x\geq100$, 
and for larger $g\beta$, $x\geq9$.}
and we estimate the error from using different fitting ranges.
Notice that to properly deal with the uncertainty due to the fit, we could run a statistical analysis as
was done in~\cite{Banuls:2013jaa},
but the simple estimate used here allows us to appreciate the relevance of reaching large $x$ values.

Finally, we obtain for the temperature dependence of the chiral condensate the improved 
results shown in Fig.~\ref{fig:final} (right), consistent with the analytical result at high
temperatures.

\section{Conclusion}
\label{sec:conclusion}

Using the massless Schwinger model as a benchmark, 
we have demonstrated that finite Matrix Product Operators can successfully describe 
thermal equilibrium states of lattice gauge theories. 
We have evaluated the thermal evolution of the chiral condensate in this model
and found good agreement with the analytical result~\cite{Sachs:1991en}
from infinite to almost zero temperature.

The high temperature region is the hardest one to capture in the continuum limit, 
maybe counterintuitively, since for typical condensed matter models it is easier to describe.
On the other hand, it is known that lattice spacing effects in the high temperature region 
can be non-negligible in conventional lattice simulations.
We have nevertheless shown that using the MPO ansatz, it is  also possible to obtain 
precise results at very small lattice spacings.

Our approach offers a systematic procedure to evaluate and control all
systematic errors in the calculation, namely bond dimension of the ansatz, 
step width of the Trotter expansion for the exponential operators, finite 
volume and continuum limit.
Although not strictly necessary, a truncation of the maximum electric flux per link can be introduced
to enhance the numerical performance. The effect of this additional cutoff parameter
is very small, but can equally be taken into account in the systematic error analysis.

All this makes the MPS  and MPO ans\"atze most valuable and promising tools 
to evaluate also other one-dimensional 
Hamiltonian systems relevant for gauge field theories. 
The most interesting open question is the extension of these techniques
to higher dimensions.

\begin{acknowledgments}
\noindent \textbf{Acknowledgments.} We are grateful to B.~Buyens, S.~K\"uhn, M.~Lubasch and
K.~Van~Acoleyen for discussions.
This work was partially funded by the EU through SIQS grant (FP7 600645),
and by the DFG Sonderforschungsbereich/Transregio SFB/TR9.
K.C. has been supported in part by the Helmholtz International Center for FAIR within the
framework of the LOEWE program launched by the State of Hessen. 
H.S. was supported by the Japan Society for the Promotion of Science for Young Scientists.
We are grateful to the computer centers of DESY Zeuthen, RZG Garching and the Center for Scientific
Computing of the Goethe-University in Frankfurt (the LOEWE system) for their computing resources
and support.  
\end{acknowledgments}


\appendix
\section{Numerical method}

For completeness, we describe here the details of the numerical method used in the paper.

To describe the thermal equilibrium state we use the MPO ansatz 
\cite{verstraete04mpdo,zwolak04mpo,pirvu10mpo},
\bea
\hat{O} =&&
\sum_{\{i_k,j_k\}}
{\rm tr} \left( M_0^{i_0j_0} 
\cdots M_{N-1}^{i_{N-1}j_{N-1}} \right ) \nn \\
&&
\hspace{15mm}
\left| i_0 \cdots i_{N-1} \right\rangle
\left\langle j_0 \cdots j_{N-1} \right|.
\label{eq:mpo}
\eea

\begin{figure}[floatfix]
\includegraphics[width=.95\columnwidth]{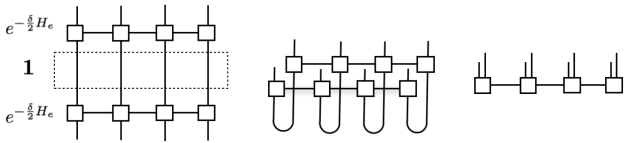}
\caption{Schematic representation of the TN operations. Left: applying the first exponential for the
first step of imaginary time evolution on the identity operator (which is an MPO with $D=1$). Center:
vectorizing this first step allows us to work with the MPS formalism. Right: after each application of
the exponential, the result is approximated by an MPS with bounded bond dimension. 
}
\label{fig:app1}
\end{figure}

The thermal density operator can be written as imaginary evolution of the identity operator~\cite{verstraete04mpdo}, 
$\rho(\beta)\propto e^{-\beta H}=e^{-\frac{\beta}{2} H} \Id e^{-\frac{\beta}{2} H}$. 
By applying a Choi isomorphism~\cite{choi}, $|i\rangle\langle j| \rightarrow |i\rangle\otimes |j\rangle$,
the operators can be vectorized. The thermal state
can thus be approximated as an MPS, by applying the imaginary time evolution operator corresponding to 
$\tilde{H} = H\otimes \Id + \Id \otimes H^{T}$ on the initial vectorized identity 
(see Fig.~\ref{fig:app1}).
As shown in Eq.~\eqref{eq:H}, the Hamiltonian contains non-commutative terms,
so we approximate the exponential operator as a sequence of MPOs, to be applied on the MPS. 
To this end, we use a 2nd-order Suzuki-Trotter expansion,
\beq
e^{-\beta \tilde{H}}\approx
\left[ e^{- \frac{\delta}{2} \tilde{H}_e} e^{- \frac{\delta}{2} \tilde{H}_z} e^{-\delta \tilde{H}_o} 
       e^{- \frac{\delta}{2} \tilde{H}_z} e^{-\frac{\delta}{2} \tilde{H}_e} \right]^M.
\label{eq:trotter}
\eeq
Each evolution step is thus approximated as the successive action of five terms.
The exponentials of the hopping terms, $H_{e(o)}=\sum_{n \ \mathrm{ even(odd)}}(\sigma_n^+ \sigma_{n+1}^-
+ h.c.)$,
have a simple exact MPO expression with maximal bond dimension $4$ \cite{pirvu10mpo},
constructed as simply the product of the individual exponentials of the mutually commuting two-body terms.
The remaining term,
\bea
   H_z
   &=& 
   \frac{\mu}{2} \sum_{n=0}^{N-1} \left[ 1+ (-1)^n \sigma_n^z \right] \nonumber \\
  && + \frac{1}{4} \sum_{n=0}^{N-2} \left[ (n+1) 
        + 2 \sum_{k=0}^n \sum_{l<k} \sigma_k^z \sigma_l^z \right] \nonumber \\
   &&+ \sum_{\substack{n=0 \\(even)}}^{N-2}  \left( 1+2\sum_{k=0}^n \sigma_k^z \right),
\eea
contains long-range terms $\sigma^z_n \sigma^z_m$,
and, although all terms commute with each other,  the product of individual exponentials
would yield a bond dimension exponentially large with the system size, $N$.

A more efficient expression for the exponential exists with bond dimension that
only scales linearly in $N$. 
We can indeed write $H_z$ as a sum of mutually commuting \emph{local} terms,
$H_z=\sum_n h_n$, where, for $n<N-1$,
\beq
h_n=\frac{\mu}{2} \left[ 1+ (-1)^n \sigma_n^z \right] +L_n^2,
\eeq
and $h_{N-1}=\frac{\mu}{2} \left[ 1+ (-1)^{N-1} \sigma_{N-1}^z \right]$,
$L_n$ being the electric flux on each link.
The exponential of $H_z$ is diagonal in the $z$ spin basis, and its value on a basis vector  
can be written as a product of the exponentials 
of each of these terms for the corresponding state.
Since the value of $L_n$, by virtue of Gauss' law, is completely determined by 
the spin content on sites $k<=n$, the factor corresponding to a given link can be 
determined from the total magnetization, $\sum_{k\leq n} \sigma_k^z$,
to the left of the corresponding site.
Such information can be encoded in the virtual index of an MPO, 
which, in a chain of length $N$, could in principle assume values $L_n\in[-N/2,N/2]$.
The exponential can thus be written exactly as an MPO determined by 
tensors $M_n$ whose only non-vanishing elements are
$(M_n^{ii})_{L_{n-1} L_n}=e^{-\delta h_n}$
for $L_{n}=L_{n-1}+\frac{1}{2}[(-1)^n+(\sigma_n^z)_{ii}]$.

Such exact expression produces an MPO with bond dimension ${\cal O}(N)$,
which is not practical for the long chains involved in our study. 
Thus, it is convenient to truncate the MPO by allowing $L_n$, i.e. the virtual index, to
assume only bounded values $|L_{n}|\leq L_{\mathrm{cut}}$, so that the maximum 
bond dimension is $2 L_{\mathrm{cut}}+1$.
This corresponds to a truncation of the physical space to only those spin configurations 
for which all links have small enough electric flux, since
the rest will be projected out when multiplying by the truncated exponential. 

Another, more economical approximation to the exponential of $H_z$
can be achieved by a 1st-order Taylor expansion, which can be 
written as an MPO with bond dimension $3$.
In this approach, the whole physical space is kept, so that no extrapolation in the $L_{\mathrm{cut}}$ 
parameter is required. In our calculation, we use both approaches.

The exponentials in \eqref{eq:trotter} involve  
$\tilde{H}_{\alpha}=H_{\alpha}\otimes \Id + \Id \otimes H_{\alpha}^{T}$ 
for each $\alpha=e,\, o,\, z$.
But since both terms in each $\tilde{H}_{\alpha}$  commute,
the corresponding exponential is just the tensor product 
of two exponentials, which can then be applied sequentially or simultaneously. 
After every factor in \eqref{eq:trotter} is written or approximated as an MPO, 
the effect of one evolution step on a certain intermediate state, vectorized as an MPS, can be 
approximated as a new MPS with the desired bond dimension.
This is achieved by a global optimization (see e.g.~\cite{verstraete08algo} for algorithmic details), in which 
the Euclidean distance 
$\epsilon = \left \| |\rho'\rangle - {O} |\rho\rangle \right \|^2$
between the new MPS, $ |\rho'\rangle$, and the result of the operator $O$ on the original one, $ |\rho\rangle$,
is minimized by successively varying each one of the tensors, and
sweeping over the chain until convergence (Fig.~\ref{fig:app1}).

\end{document}